\documentstyle[12pt]{article}

\makeatletter
\@addtoreset{equation}{section}

\newcommand{\nn}{\nonumber}
\newcommand{\vs}[1]{\vspace*{#1}}
\newcommand{\hs}[1]{\hspace*{#1}}
\newcommand{\tr}{\mathop{\rm tr}}
\newcommand{\p}{\partial}
\newcommand{\Half}{\frac12}
\newcommand{\unit}{\hbox to 3.8pt{\hskip1.3pt \vrule height 7.4pt
    width .4pt \hskip.7pt \vrule height 7.85pt width .4pt \kern-2.4pt
    \hrulefill \kern-3pt \raise 3.7pt\hbox{\char'40}}}

\def\href#1#2{#2}
\newcommand{\ket}[1]{|{#1}\rangle}
\newcommand{\bra}[1]{\langle{#1}|}

\begin{document}

\begin{titlepage}

\bigskip

\title{
\vs{-10mm}
\hfill\parbox{4cm}{
{\normalsize\tt hep-th/0105311}\\[-5mm]
{\normalsize \tt CALT-68-2331}\\[-5mm]
{\normalsize \tt CITUSC/01-019}\\[-5mm]
{\normalsize \tt NSF-ITP-01-42}
}
\\[50pt]
~\\Seiberg-Witten Transforms of Noncommutative Solitons\\
~}
\author{
Koji {\sc Hashimoto}\thanks{\tt
koji@itp.ucsb.edu}$\hspace{2mm}{}^a$
and
Hirosi {\sc Ooguri}$\thanks{\tt
ooguri@theory.caltech.edu}\hspace{2mm}{}^{a,b}$
 \\ \\ [8pt]
${}^a${\it Institute for Theoretical Physics,
University of California, }\\
{\it Santa Barbara, CA 93106-4030}\\
${}^b${\it California Institute of Technology 452-48,
Pasadena, CA 91125}\\ }

\date{}

\maketitle
\thispagestyle{empty}

\begin{abstract}
\normalsize\noindent
We evaluate the Seiberg-Witten map for solitons
and instantons in noncommutative gauge theories
in various dimensions. We show that solitons
constructed using the projection operators
have delta-function supports when expressed in
the commutative variables. This gives a precise
identification of the moduli of these solutions
as locations of branes. On the other hand, an instanton
solution in four dimensions allows deformation
away from the projection operator construction.
We evaluate the Seiberg-Witten transform of the
$U(2)$ instanton and show that it has a finite size
determined by the noncommutative scale and by the deformation
parameter $\rho$. For large $\rho$,
the profile of the D$0$-brane density of the instanton
agrees surprisingly well with that of the BPST instanton on
commutative space.

\end{abstract}

\end{titlepage}


\section{Introduction}

Noncommutative gauge theories can be realized by considering
branes in string theory with a constant NS-NS two-form
field \cite{Connes:1998cr}. It is described by noncommutative gauge
fields $\hat{A}_i$ on a noncommutative space whose coordinates
obey the commutation relation,
\begin{eqnarray} [ \hat{x}^i, \hat{x}^j ] = i\theta^{ij}.
\label{noncomm}
\end{eqnarray}
One of the remarkable features of these theories is that there
is a universal way to construct a large class of classical
solutions \cite{NS} -- \cite{latest}. In particular, in
2 dimensions,
all solutions to the noncommutative Yang-Mills equations
with gauge group $U(N)$
are classified in \cite{GN3}, and it was shown
that they take the form
\begin{eqnarray} X^i = U \hat{x}^i
U^\dagger + \sum_{a=1}^m \lambda_a^i
|a \rangle
\langle a |~~~~~(i=1,2),
\label{2dsolution}\end{eqnarray}
where
\begin{eqnarray}
X^i = \hat{x}^i - \theta^{ij} \hat{A}_j(\hat{x})
\label{matrix}
\end{eqnarray}
are operators acting on the Hilbert space
${\cal H}$, which is the Fock space
of (\ref{noncomm}) times ${\bf C}^N$,
$\{ | a \rangle \}_{a=1\cdots m}$ is an $m$-dimensional
subspace of ${\cal H}$, and $U$ is the associated
shift operator obeying
\begin{eqnarray}
 U^\dagger U = 1, ~~ UU^\dagger = 1 - \sum_{a=1}^m
 |a \rangle \langle a | .
\end{eqnarray}
Thus the solutions are parameterized by the rank $m$ of the projection
operator $1 - UU^\dagger$, the rank $N$ of the gauge
group\footnote{It may not be evident in the expression
(\ref{2dsolution})
that the rank $N$ of gauge group is a parameter of
the solution invariant under the $U(\infty)$ gauge symmetry.
To see that there is a gauge invariant definition of
$N$, we point out the formula derived in \cite{OO,Mukhi,Jeremy}:
\begin{eqnarray}
 {\rm Tr} \left[ {\rm Pf}\left([X^i, X^j]\right)
e^{ik\cdot X} \right] = N ~\delta(k).
\end{eqnarray}
This holds as far as the gauge field $\hat{A}_i(x)$
has a compact support when it is expressed in terms
of commutative
variables via the Seiberg-Witten map.
One may
also be able to show that $N$ is gauge invariant
by using the more precise definition  of the
$U(\infty)$ group recently given in \cite{harveydef}.},
and the $2m$ moduli parameters $\lambda_a^i$.
These solitons are interpreted as D0-branes on D2-branes
with $m$ and $N$ being the D0 and D2 charges respectively.
There have been evidences suggesting that $\lambda_a^i$ correspond to
the locations of the D0-branes \cite{Gopak,GN3,Aki}. In this paper
we will confirm this interpretation using the Seiberg-Witten map.
In higher dimensions, a complete classification of solutions
has not been carried out, although some special solutions
are known such as instanton solutions in four dimensions,
which can be interpreted as D0-branes on D4-branes \cite{NS} -- \cite{
sonota2}\cite{sonota3,sonota4,sonota5, KF}.
These higher dimensional solutions do not necessarily
take the form (\ref{2dsolution}).

In \cite{Seiberg}, it was shown that there are two equivalent
descriptions of the theory, one in terms of ordinary
gauge fields $A_i$ on a commutative space and another in terms
of noncommutative gauge fields $\hat{A}_i$ on a noncommutative
space. The map between $A_i$ and $\hat{A}_i$ is called the
Seiberg-Witten map. In \cite{OO,Mukhi,Jeremy}, an explicit expression
for the Seiberg-Witten map was found for the $U(1)$ part of
the field strength, by studying the coupling of
the gauge field to the Ramond-Ramond potentials of closed
string in the bulk\footnote{There has also been an
  approach \cite{Okuyama,Jurco1,Jurco2, Jurco3} 
to express the Seiberg-Witten map
  using the Kontsevich formal map \cite{Kontsevich,Felder}.}.
The expression was conjectured earlier in \cite{Liu}.
It was proven in \cite{OO}
that it indeed satisfies the conditions
for the Seiberg-Witten map without relying on the string theory origin
of the expression. In this paper, we evaluate the
Seiberg-Witten map  for the noncommutative soliton solutions
in the above paragraph and express them in terms of the commutative
variables.

In two dimensions, where a solution always takes the form
(\ref{2dsolution}), we find that the $U(1)$ part of
the commutative field strength has a delta-function support
at $x^i = \lambda_a^i$. This confirms the earlier observation
that the moduli $\lambda_a^i$ should be regarded as positions
of D0-branes on the D2-branes. It is interesting to
note that $\lambda_a^i$ are commutative parameters even though they
are describing the locations of the noncommutative solitons.
A natural explanation for this is that the coordinates $x^i$ of the
commutative variables $A_i(x)$ should be considered as
the closed string coordinates, which are commutative,
since the Seiberg-Witten
map we use was derived from the study of the coupling of the gauge
theory to the Ramond-Ramond potentials in the bulk.
It is rather surprising that, whether the gauge group is
Abelian or non-Abelian, all the solutions in 2 dimensions
are singular when expressed in terms of the commutative variables
$A_i(x)$. The fact that there is no moduli which change the
size of the solitons has been known from the analysis of the
massless modes of the open string connecting D0-branes and
D2-branes, but one may have expected that the soliton has
a fixed size set by the noncommutative parameter $\theta^{ij}$.
This turned out not to be the case for these solutions.
There are various other solutions, describing
branes intersecting with each other with arbitrary angles,
which can be expressed
in the form (\ref{2dsolution}), and they all have
delta-function singularities after the Seiberg-Witten
transform.

On the other hand, solutions in higher dimensions
are not necessarily of the form (\ref{2dsolution})
and therefore can have a finite size after the Seiberg-Witten
transform. We examine in detail the $U(2)$ instanton
constructed in \cite{KF}.
The solution contains an extra modulus
$\rho$, which in the commutative limit $\theta \rightarrow 0$
reduces to the size of the instanton. We evaluate the
Seiberg-Witten transform of this solution in the two limit,
$\rho \ll \sqrt{\theta}$ and $\sqrt{\theta} \ll \rho$. When $\rho=0$,
the instanton solution is of the form (\ref{2dsolution})
and has a delta-function singularity when expressed in
the commutative variables. We find that, as soon as we
turn on a small amount of $\rho$, the solution gets a non-zero support
of the size $\sim \sqrt{\theta}$. We also see that the
delta-function singularity is modified by $\rho$.
On the other hand, for $\sqrt{\theta} \ll \rho$, we find
that the delta-function singularity is completely resolved
and that the solution has a smooth profile, which,
for the first two terms in
the $1/\rho$ expansion,
precisely agrees with that of the BPST instanton
on commutative space.

This paper is organized as follows. In Sec.\ 2, we review the
construction of the Seiberg-Witten map derived in
\cite{OO,Mukhi,Jeremy}.  In Sec.\ 3, we evaluate Seiberg-Witten
transform of the noncommutative  solitons in (2+1) dimensions, which
take the form (\ref{2dsolution}).  Other examples including
intersecting branes and fluxons are discussed in Sec.\ 4. In Sec.\ 5,
we study the Seiberg-Witten transform of the $U(2)$ noncommutative
instanton solution and show how the delta-function singularity is
resolved. We will close this paper with discussions of
our results in Sec.\ 6.
In Appendices, we derive some of the formulae used in this
paper and give some details of the computation in Sec.\ 5. 


\section{Seiberg-Witten Map}

In \cite{OO,Mukhi,Jeremy}, an exact and explicit form of the
Seiberg-Witten map for the $U(1)$ part of the field strength
was obtained from string theory computation of the
coupling between the noncommutative gauge theory on the branes
and the Ramond-Ramond potentials in the bulk. For a gauge theory
with $2n$ noncommutative dimensions, the map from
the field strength in the noncommutative variables $\hat{A}_i$
\begin{eqnarray}
\hat{F}_{ij} = \partial_i \hat{A}_j - \partial_j \hat{A}_i
+ i \hat{A}_i * \hat{A}_j - i\hat{A}_j * \hat{A}_i
\end{eqnarray}
to the field strength
$F_{ij} = \partial_i A_j - \partial_j A_i$ of the commutative
variables $A_i$ is given\footnote{
In this paper, we choose the sign of the noncommutative
parameter $\theta^{ij}$ as in (\ref{noncomm}).
To use the convention in \cite{OO},
one can simply make the substitution $\theta^{ij} \rightarrow -
\theta^{ij}$ in the following.}
 in the Fourier transformed form by
\begin{eqnarray}
\lefteqn{  F_{ij}(k) - \theta_{ij}^{-1} \delta(k)}
\nn\\
&&=\frac{1}{{\rm Pf}(\theta)}
\int \! dx *
\left[
  e^{ik\cdot X}
\left(\theta + \theta \hat{f} \theta\right)_{ij}^{n-1}
P
\exp
\left(
  i\int^1_0\hat{A}_i(\hat{x}+l\tau)l^id\tau
\right)
\right],
\label{SWmap}
\end{eqnarray}
where
 \begin{eqnarray}
&& \left(\theta + \theta \hat{f} \theta\right)_{ij}^{n-1}
= {1 \over 2^{n-1}(n-1)!} \epsilon_{iji_1i_2\cdots i_{2n-2}}
\nn\\
&&~~~~~~\times
 \int_0^1 d\tau_1 \left(\theta + \theta
\hat{F}(\hat{x}+l\tau_1)\theta\right)^{i_1i_2}
\cdots
\int_0^1 d\tau_{n-1} \left(\theta + \theta
\hat{F}(\hat{x}+l\tau_{n-1})\theta\right)^{i_{2n-3}i_{2n-2}}
\end{eqnarray}
In particular, for $n=1$ and $2$, we have
\begin{eqnarray}
  \left(\theta + \theta \hat{f} \theta\right)_{ij}^{n-1}
=\left\{
  \begin{array}{ll}
\epsilon_{ij} & (n=1), \\
\Half \epsilon_{ijkl}
{\displaystyle \int}_0^1d\tau
\left(
  \theta+\theta\hat{F}(\hat{x}+l\tau)\theta
\right)^{kl} & (n=2).
  \end{array}
\right.
\end{eqnarray}
This expression involves the open Wilson line, which
is a basic building block of observables in
noncommutative gauge theory \cite{Ishibashi:2000hs,
Das:2000md,Aki, Bak}.
In order to actually evaluate the Seiberg-Witten map,
it is useful to express it using the variable $X^i$ defined
by (\ref{matrix}). For $n=1$, the Seiberg-Witten map is
given by
\begin{eqnarray}
{}F_{12}(k) - \theta_{12}^{-1} \delta(k)
={\rm Tr} e^{ik\cdot X},
\label{sw1}
\end{eqnarray}
and for $n=2$ by
\begin{eqnarray}
 {} F_{ij}(k) - \theta_{ij}^{-1} \delta(k)
=-\frac{i}{2}\epsilon_{ijkl}{\rm Tr}
 \left([X^k, X^l]e^{ik\cdot X}\right).
\label{sw2}
\end{eqnarray}

When the noncommutative gauge theory is realized on
D$p$-branes,
the field strength $F_{ij}(k)$ of the commutative variables
$A_i(x)$  can be regarded as the D$(p-2)$-brane
density on the D$p$-branes. This was how the expression (\ref{SWmap})
was found in \cite{OO,Mukhi,Jeremy}. 
In the following, we will find it useful to
consider lower brane densities also. The D$(p-2s)$-brane
density on the D$p$-branes is given by
\begin{eqnarray}
 J_{i_1 \cdots i_{p-2s}} &\sim &
\epsilon_{i_1 \cdots i_{p-2s} j_1 \cdots j_{2s}}
  \int_0^1 d\tau_1 \int_{\tau_1}^1 d\tau_2
\cdots \int_{\tau_{s-2}}^1 d\tau_{s-1}
\\
&&~~\times {\rm Tr}\left(
 [X^{j_1},X^{j_2}] e^{i\tau_1 k\cdot X}
[X^{j_3},X^{j_4}] e^{i(\tau_2-\tau_1) k\cdot X}
\cdots [X^{j_{2s-1}},X^{j_{2s}}]
e^{i(1-\tau_{s-1})k\cdot X}\right).
\nn
\label{generalcharge}
\end{eqnarray}


\section{Solitons in 2+1 dimensions}

In \cite{GN3}, all static classical solutions
to the noncommutative Yang-Mills theory in $(2+1)$
dimensions are classified.
They take the form
\begin{eqnarray} X^i = U \hat{x}^i U^\dagger
+ \sum_{a=1}^m \lambda_a^i
|a \rangle
\langle a |~~~~~(i=1,2),
\label{generalsolution}
\end{eqnarray}
where
$\{ | a \rangle \}_{a=1\cdots m}$ is an $m$-dimensional
subspace of
the Fock space of (\ref{noncomm}) times ${\bf C}^N$,
$\lambda_a^i$'s are arbitrary constant parameters,
and $U$ is the associated
shift operator obeying
\begin{eqnarray}
 U^\dagger U = 1, ~~ UU^\dagger = 1 - \sum_{a=1}^m
 |a \rangle \langle a | .
\end{eqnarray}
It is straightforward to compute the Seiberg-Witten
transform of this solution\footnote{This
is essentially the same as the computation of the Wilson line
observables in the soliton background discussed
in \cite{GN3,Aki}. Here we are reinterpreting
it as an evaluation of the Seiberg-Witten map.}.

Substituting (\ref{generalsolution}) into the
Seiberg-Witten map (\ref{sw1}), we find
\begin{eqnarray}
\tr e^{ik\cdot X}
&=& \tr
\left[
  U e^{ik\cdot\hat{x}} U^\dagger
+ \sum_{a=0}^{m-1}e^{ik_i\lambda_a^i} \ket{a}\bra{a}
\right] \nn\\
&=& {\rm tr}e^{ik\cdot \hat{x}} +
\sum_{a=0}^{m-1}e^{ik_i\lambda_a^i} \langle a | a\rangle
\nn\\
&=&\frac{1}{\theta} \delta(k)
+ \sum_{a=0}^{m-1}e^{ik_i\lambda_a^i}.
\end{eqnarray}
Here in the first equality we have used the following identity
\begin{eqnarray}
  e^{iUk\cdot\hat{x}U^\dagger} = U e^{ik\cdot\hat{x}} U^\dagger + 1 -
  UU^\dagger.
\label{important}
\end{eqnarray}
The field strength expressed in the commutative variables
is then\footnote{Note that, since
$\theta^{ij}$ is antisymmetric, $\theta_{12}^{-1} =
-1/\theta^{12}$.}
\begin{eqnarray}
  {}F_{12}(k) = \sum_{a=0}^{m-1}e^{ik_i\lambda_a^i}.
\end{eqnarray}
By taking the Fourier transform of this, we find
\begin{eqnarray}
  {}F_{12}(x) = \sum_{a=0}^{m-1} \delta(x-\lambda_a).
\end{eqnarray}
We see that the solution has delta-function supports
at $x=\lambda_a$ ($a=0,\cdots,m-1$). This gives a
precise interpretation of the moduli
$\lambda_a$ as representing the locations of the
soliton, confirming the observations in
\cite{Gopak,GN3,Aki}.

There is an obvious generalization of this construction
to higher dimensions. Let us assume that $\theta^{12},
\theta^{34},\cdots, \theta^{i_{2n-1}i_{2n}} \neq 0$
and other $=0$ so that we have a direct product of
$n$ Fock spaces. We can then consider a solution,
\begin{eqnarray}
  X^i = U x^i U^\dagger,~~~~(i=1,\cdots,2n).
\label{obvious}
\end{eqnarray}
Here we set all the moduli $\lambda = 0$ for simplicity,
and $U$ is a shift operator of rank $m$.
The Seiberg-Witten map in $2n$ dimensions is
\begin{eqnarray}
  {}F_{ij}(k)-  \theta_{ij}^{-1}\delta(k)
& = &
-{i^{n-1}  \over 2^{n-1} (n-1)} \epsilon_{ijs_1\cdots s_{2n-2}}
  \int_0^1 d\tau_1 \cdots \int_{\tau_{n-3}}^1 d\tau_{n-2}
\nn\\
&&\times
{\rm tr}
\left( [X^{s_1},X^{s_2}] e^{i\tau_1 k\cdot X}
   \cdots [X^{s_{2n-3}},X^{s_{2n-2}}]
e^{i(1-\tau_{n-2}) k\cdot X} \right) .
\label{one}
\end{eqnarray}
Using (\ref{important}) and
\begin{eqnarray}
 [X^i, X^j] = i U\theta^{ij} U^\dagger,
\end{eqnarray}
one finds that the right-hand side of (\ref{one}) is
$-\theta^{-1}_{ij}\delta(k)$, and therefore
\begin{eqnarray}
{}F_{ij}(k) = 0.
\label{zero}
\end{eqnarray}
Similarly one can show that the soliton
does not give a nontrivial contribution to the
the D$2p$-brane density for all $p\geq 1$.
The only non-vanishing one is the D0-brane density,
which is given by
\begin{eqnarray}
J(k)&=& {\rm Tr}\left( e^{ik\cdot X} \right)
\nn\\
 &=& {\rm Tr} \left( U e^{ik\cdot \hat{x}} U^\dagger +1 - UU^\dagger
 \right)
\nn\\
& =& {1\over {\rm Pf}(\theta)} \delta(k) + m.
\end{eqnarray}
The Fourier transform of this gives
\begin{eqnarray}
J(x) = {1\over {\rm Pf}(\theta)} + m \delta(x).
\end{eqnarray}
The first term represents the background D0-brane
charge in the presence of the constant $B$ field
and the second term corresponds to the $m$ D0-branes
described by the soliton solution (\ref{obvious}).
This soliton therefore describes $m$ D$0$-branes without
higher brane charges.

One may be puzzled by that fact that the solution (\ref{obvious})
of the noncommutative $U(1)$ gauge theory describes D$0$-branes
even though the field strength
$F_{ij}$ of this solution is identically equal to zero!
Such a bizarre behavior is not unexpected for solutions
with delta-function singularities. To illustrate the point,
let us imagine that $F_{ij}$ has the following
configuration,
\begin{eqnarray}
{}F_{ij} \sim \epsilon^{-2} \exp\left(-{x^2\over \epsilon^2}\right).
\label{two}
\end{eqnarray}
In this case,
\begin{eqnarray}
 \epsilon^{i_1\cdots i_{2n}} F_{i_1i_2} \cdots F_{i_{2n-1}i_{2n}}
 \sim \epsilon^{-2n} \exp\left( - n{x^2 \over \epsilon^2}\right).
\label{three}
\end{eqnarray}
In the limit $\epsilon \rightarrow 0$, the field strength
vanishes $F_{ij} \rightarrow 0$, but $F^n$ becomes
proportional to $\delta(x)$.

If we embed the solution (\ref{obvious}) to the
$U(N)$ gauge theory, it is possible to deform it away from the form
(\ref{obvious}). In Sec.\ 5, we study the $U(2)$ instanton solution in
four dimensions,  for which an explicit expression is known
\cite{KF}. We find that the Seiberg-Witten transform of the instanton
acquires a finite size as soon as we turn on the deformation, and the
size is set by the noncommutative scale $\theta$ and the deformation
parameter $\rho$. We also show the $U(1)$ part of the field strength
becomes non-zero after the deformation.


\section{Intersecting Branes}

Noncommutative soliton solutions representing orthogonally
intersecting branes have been constructed in literature \cite{berg,
  tseng}. In this section, we generalize these constructions by
allowing arbitrary angles and evaluate their Seiberg-Witten
transforms.


\subsection{D2-branes orthogonally intersecting on D4-brane}

As a warm-up, let us consider D$2$-branes orthogonally
intersecting on a D$4$-brane worldvolume. It can be obtained
by reinterpreting the tachyon configuration studied
in \cite{berg} as a gauge field configuration on the
D$4$-brane:
\begin{eqnarray}
  X^{1,2} = U \hat{x}^{1,2} U^\dagger \otimes \unit,
\quad
  X^{3,4} = \unit \otimes V \hat{x}^{3,4} V^\dagger.
\label{orthogonal}
\end{eqnarray}
Here we introduced noncommutativity as $\theta^{12}, \theta^{34}\neq
0$,  and so we have a direct product of the two Fock spaces.
The operator $V$ is the same as $U$ except that $V$ acts on the 
second Fock space of $\hat{x}^3$ and $\hat{x}^4$:
\begin{eqnarray}
  U \equiv \sum_n \ket{n+m} \bra{n} \otimes \unit,\quad
  V \equiv \unit \otimes \sum_n \ket{n+l} \bra{n}.
\end{eqnarray}
The above solution represents the brane configuration in which
$m$ D2-branes localized at the origin of
the $x^1$-$x^2$ plane are intersecting with
$l$ D2-branes localized at the origin of the $x^3$-$x^4$ plane.
This geometrical interpretation is confirmed by
evaluating the Seiberg-Witten map (\ref{sw2}) for four noncommutative
dimensions:
\begin{eqnarray}
  {}F_{12}(x) = m\delta(x^1)\delta(x^2),
\quad
  {}F_{34}(x) = l \delta(x^3)\delta(x^4),
\quad
{\rm others}=0.
\label{intersectinggauge}
\end{eqnarray}
It is also interesting to calculate the D$0$-brane density
using (\ref{generalcharge}) with $p=4$, $s=2$:
\begin{eqnarray}
J(k) &=&  {\rm Tr} (e^{ik\cdot X}) \nn \\
& =& {\rm Tr}
\left[
   U\exp(ik_1\hat{x}^1+ik_2\hat{x}^2)U^\dagger
    \otimes    V\exp(ik_3\hat{x}^3+ik_4\hat{x}^4)V^\dagger
\right.
\nn\\
&& \hs{10mm}\left.
 +\sum_{a=0}^{m-1}  \ket{a}\bra{a}
 \otimes     V\exp(ik_3\hat{x}^3+ik_4\hat{x}^4)V^\dagger
\right.
\nn\\
&& \hs{10mm}\left.
+U\exp(ik_1\hat{x}^1+ik_2\hat{x}^2)U^\dagger
\otimes
\sum_{a=0}^{l-1}  \ket{a}\bra{a}
 +\sum_{a=0}^{m-1}  \ket{a}\bra{a}
\otimes
\sum_{a=0}^{l-1}  \ket{a}\bra{a}
\right]
\nn\\
&=&
{1\over \theta^{12}\theta^{34}}\delta^4(k)
+ {m\over \theta^{34}} \delta(k_3)\delta(k_4)
+ {l \over \theta^{12}} \delta(k_1)\delta(k_2)
+ ml.
\end{eqnarray}
After the Fourier transformation, we obtain
\begin{eqnarray}
  J(x) = {1\over \theta^{12}\theta^{34}}
 + {m\over \theta^{34}} \delta(x^1)\delta(x^2)
 + {l\over \theta^{12}}\delta(x^3)\delta(x^4) + ml \delta^4(x).
\label{d0cha}
\end{eqnarray}
It is interesting to note that, using (\ref{intersectinggauge}),
this can be expressed as
\begin{eqnarray}
  J(x) = {1\over 8} \epsilon^{ijkl}
\left(
 {}F_{ij}(x) -\theta^{-1}_{ij}
\right)\left(
 {}F_{kl}(x) -\theta^{-1}_{kl}
\right).
\end{eqnarray}
Such a relation between the D$0$-brane charge density
$J(x)$ and the field strength $F_{ij}$ holds in the
leading order in the standard $\alpha'$ expansion of string
theory computation, but it is expected to receive large corrections
in the Seiberg-Witten limit. In fact, in the more elaborate
examples discussed  below, such a relation does not hold.


\subsection{Intersection with arbitrary angles}

We can introduce an arbitrary angle to the solution
(\ref{orthogonal}) by deforming it as follows,
\begin{eqnarray}
&&  X^{1,2} = U \hat{x}^{1,2} U^\dagger \otimes \unit
  +\sum_{a=0}^{m-1}  \ket{a}\bra{a}\otimes
\lambda_a^{1,2}(\hat{x}^3,\hat{x}^4),
\label{additionalmoduli2}\\
&&  X^{3,4} = \unit \otimes  \hat{x}^{3,4} ,
\label{additionalmoduli}
\end{eqnarray}
where $\lambda$'s are functions
of $\hat{x}^3$ and $\hat{x}^4$.
Here we have set $l=0$ so that the configuration does not include
localized D0-branes (see the last term in (\ref{d0cha}).)
Substituting this into the equation of motion,
\begin{eqnarray}
   [X_i, [X^i, X^j]] = 0,
\end{eqnarray}
we find that $\lambda$'s have to be linear functions,
\begin{eqnarray}
  \lambda_a^i(\hat{x}^3, \hat{x}^4)
= \alpha_a^i + \beta_a^i \hat{x}^3 + \gamma_a^i \hat{x}^4
\end{eqnarray}
where $\alpha$, $\beta$ and $\gamma$ are constant parameters, and
$i=1,2$.

We can regard $\lambda_a$'s as representing the configurations
of the D2-branes. To confirm this interpretation, we evaluate
the Seiberg-Witten map (\ref{sw2}).
\begin{eqnarray}
{}F_{12}(x) =   \sum_{a=0}^{m-1}\delta_a(x),\quad
&&
{}F_{34}(x) =   \sum_{a=0}^{m-1}(\beta_a^1\gamma_a^2
-\beta_a^2\gamma_a^1)\delta_a(x),
\nn\\
{}F_{13}(x)= -\sum_{a=0}^{m-1}\beta_a^2\delta_a(x),\quad
&&
{}F_{23}(x)= \sum_{a=0}^{m-1}\beta_a^1\delta_a(x),
\nn\\
{}F_{14}(x)= -\sum_{a=0}^{m-1}\gamma_a^2\delta_a(x),\quad
&&
{}F_{24}(x)= \sum_{a=0}^{m-1}\gamma_a^1\delta_a(x),
\label{bianchiok}
\end{eqnarray}
where
\begin{eqnarray}
\delta_a(x)\equiv
\delta\left(
x^1-\lambda_a^1(x^3, x^4)
\right)
\delta\left(
x^2-\lambda_a^2(x^3, x^4)
\right).
\end{eqnarray}
Therefore the D2-branes are located as expected.
It is also useful to point out that (\ref{bianchiok})
satisfies the Bianchi identity, $\partial_{[i,}
{}F_{j,k]}=0$. For example,
\begin{eqnarray}
 \partial_{[1,} F_{2,3]}
&=& \sum_a \left( \partial_3 +
\beta_a^1 \partial_1 +
 \beta_a^2 \partial_2\right) \delta(x^1-\lambda_a^1)
\delta(x^2-\lambda_a^2)
\nn\\
&=& 0,
\end{eqnarray}
consistently with the general proof
in \cite{OO}.

The D$0$-brane density for this solution is
\begin{eqnarray}
  J(k)
&=& {\rm Tr}
\left[
   U\exp(ik_1\hat{x}^1+ik_2\hat{x}^2)U^\dagger
    \otimes    \exp(ik_3\hat{x}^3+ik_4\hat{x}^4)
\right.
\nn\\
&& \hs{10mm}
\left.
+\sum_{a=0}^{m-1}
\ket{a}\bra{a}
\otimes
\exp(ik_1\lambda_a^1
+ik_2\lambda_a^2+ik_3\hat{x}^3+ik_4\hat{x}^4)\right]
\nn\\
& =& \frac{1}{{\rm Pf}(\theta)}
\delta^4(k) + \frac{1}{\theta^{34}}
\sum_a e^{ik_a \alpha_a^1+ik_2\alpha_a^2}
\delta(k_1 \beta_a^1+k_2 \beta_a^2 + k_3)
\delta(k_1 \gamma_a^1+k_2 \gamma_a^2 + k_4).
\end{eqnarray}
After performing the Fourier transformation,
we obtain
\begin{eqnarray}
  J(x) = \frac{1}{{\rm Pf}(\theta)} +
\frac1{\theta^{34}}\sum_{a=0}^{m-1}
\delta_a(x).
\end{eqnarray}
As before, the first term shows the uniform distribution of the
D0-branes in the D4-brane. The second term indicates the D0-branes
bound in the D2-branes located at the place where the delta-functions
specify. There is no localized D0-brane in this case.

We have shown that it is possible to introduce moduli
to the intersecting brane solutions as in 
(\ref{additionalmoduli2}) and (\ref{additionalmoduli})
to describe configurations of branes with arbitrary angles.
We can generalize this further by introducing additional
moduli
as
\begin{eqnarray}
 && X^{1,2} = U \hat{x}^{1,2} U^\dagger\otimes \unit
  +\sum_{a=0}^{m-1} \ket{a}\bra{a}
\otimes \lambda_a^{1,2}(\hat{x}^3,\hat{x}^4)
+ \sum_{a,b} \zeta_{ab}^{1,2}\ket{a}\bra{a}\otimes\ket{b}\bra{b} ,
\label{add1}\\
 && X^{3,4} = \unit \otimes
V \hat{x}^{3,4} V^\dagger
  +\sum_{b=0}^{l-1} \lambda_b^{3,4}(\hat{x}^1, \hat{x}^2)
\otimes \ket{a}\bra{a}
+ \sum_{a,b} \zeta_{ab}^{3,4}\ket{a}\bra{a}\otimes\ket{b}\bra{b}.
\label{add2}
\end{eqnarray}
The Seiberg-Witten map gives
\begin{eqnarray}
&&  F_{12}(x)=
\sum_{a=0}^{m-1}\delta\left(x^1-\lambda_a^1(x^3, x^4)\right)
\delta\left(x^2-\lambda_a^2(x^3, x^4)\right)\nn\\
&&\hspace{30mm}
+\sum_{b=0}^{l-1}(\beta_b^3\gamma_b^4
-\beta_b^4\gamma_b^3)\delta\left(x^3-\lambda_a^3(x^1, x^2)\right)
\delta\left(x^4-\lambda_a^4(x^1, x^2)\right),
\end{eqnarray}
and similar expressions for the other components of the gauge field
strength. Note that the number of the D0-branes is $ml$ whereas the
number of the D2-branes is $m+l$.
The D0-brane density is given by
\begin{eqnarray}
  J(x) = \frac{1}{{\rm Pf}(\theta)}
+\frac{1}{\theta}
\sum_{a=0}^{m-1}\delta\left(x^1-\lambda_a^1(x^3, x^4)\right)
\delta\left(x^2-\lambda_a^2(x^3, x^4)\right)
\nn\\
+
\frac{1}{\theta^{12}}
\sum_{b=0}^{l-1}\delta\left(x^3-\lambda_a^3(x^1, x^2)\right)
\delta\left(x^4-\lambda_a^4(x^1, x^2)\right)
\nn\\
+\sum_{a,b}\Pi_{i=1}^4\delta(x^i-\zeta_{ab}^i).
\end{eqnarray}
The last term shows the localized D0-branes scattered in the
D4-brane.

It is straightforward to include the scalar field in this
construction and allow the D2-branes and the D0-branes to 
move away from the D4-brane, as discussed in \cite{Gopak}.

\subsection{D1-branes intersecting with D3-brane}

The solutions discussed so far are all non-BPS
and unstable. One of the interesting BPS noncommutative solutions
is the fluxon solution studied in \cite{GN2, GN3, Tera, KH}.
If we turn on the noncommutativity only along the $x^1$-$x^2$ plane,
the solution representing D1-branes piercing a D3-brane is
\begin{eqnarray}
  X^i = U \hat{x}^i U^\dagger + \sum_{a=0}^{m-1} \ket{a}\bra{a}
  \lambda_a^i,\quad \nn\\
A_3=0, \quad
\hat{\Phi} = \frac{1}{\theta^{12}}
\sum_{a=0}^{m-1} (x_3 - \zeta_a)\ket{a}\bra{a}.
\label{fluxonsol}
\end{eqnarray}
Note that $i=1,2$ and there is no noncommutativity along $x^3$.
The above solution satisfies the BPS equations in noncommutative
Yang-Mills theory on the D3-brane,
\begin{eqnarray}
&& - \p_3 \hat{\Phi} = B_3
\equiv \frac{-i}{(\theta^{12})^2}
\left([X^1,   X^2]-i\theta^{12}
\right), \nn\\
&&i[X^1, \hat{\Phi}]/\theta^{12} = \hat{B}_2 ,\quad
-i[X^2, \hat{\Phi}]/\theta^{12} = \hat{B}_1 .
\end{eqnarray}
The last two equations are trivially satisfied
since both sides of the two equations vanish.

The D1-brane current density of this solution is
\begin{eqnarray}
\lefteqn{  {\rm Tr} \exp\left(
ik_1X^1 + ik_2X^2 + ik_\Phi \Phi
\right)}\nn\\
&& = \frac{1}{\theta^{12}}\delta(k_1)\delta(k_2)+\sum_a
\exp\left(ik_1\lambda_a^1 + ik_2 \lambda_a^2+ik_\Phi
\frac{1}{\theta^{12}}
(x_3-\zeta_a)\right),
\end{eqnarray}
where note that we have introduced a
transverse momentum $k_\Phi$ coupled to $\Phi$.
The Fourier transform of this expression is
\begin{eqnarray}
  J(x) &=& \int dk_1 dk_2 dk_\Phi \;
e^{-ik_1 x^1 - ik_2 x^2 - ik_\Phi\Phi}
  J(k) \nn\\
&=& \frac{1}{\theta^{12}}\delta(\Phi) +
\sum_a\delta(x^1-\lambda_a^1)\delta(x^2-\lambda_a^2)\delta(\Phi
-(x_3-\zeta_a)/\theta^{12}).
\end{eqnarray}
The first term shows the D1-branes uniformly distributed on
the D3-brane as a result
of the background B-field $B_{12}$.  The second term shows the
D1-branes intersecting with the D3-brane. We note that
the intersection angle depends on
$\theta$, as expected for the BPS solution. 
The intersection point is located at $(\lambda_a^1, \lambda_a^2,
\zeta_a)$ on the worldvolume of the D3-brane.

It is easy to generalize this solution to various other cases,
$e.g.$, infinite number of D1-branes piercing \cite{KH},  introducing
another transverse scalar field in such a way that the D1-brane is
completely apart from the D3-brane \cite{Tera}, and non-BPS
deformation by changing the tilt of the D1-brane \cite{GN3}. The
Seiberg-Witten transforms of these solutions confirm the known
interpretations of these solitons and their moduli.


\section{Instantons and Resolution of the Delta Function
  Singularities} 

We have found that solutions constructed using
projection operators have delta-function singularities.
In this section, we will study how these singularities
are resolved in the case of
the $U(2)$ instanton solution on the four dimensional noncommutative
space with a single scale modulus $\rho$.

{}For definiteness, we assume that the noncommutative
parameter $\theta^{ij}$ is anti-self-dual and set
\begin{eqnarray}
 \theta^{34} = - \theta^{12} = \theta>0,~~~{\rm other} = 0.
\label{seven}
\end{eqnarray}
Given this, there is a distinction between self-dual and anti
self-dual solutions, constructed in \cite{NS, hsy} and in \cite{KF}
respectively.
In this section, we examine the anti-self-dual solution
of \cite{KF} since it can be regarded as a deformation of a solution
of the form (\ref{2dsolution}) embedded in the $U(2)$ theory,
as we will see explicitly in (\ref{four}) and (\ref{five}).

Let us review the construction of the anti-self-dual solution in
\cite{KF}. To simplify the computations in the following,
we rescale the coordinates $\hat{x}^i$ so that the noncommutative
scale is set as $\theta=1$. Whenever necessary, we can
restore $\theta$ by a simple dimensional analysis.
It is useful to combine the coordinates into
the form of the creation and annihilation operators
\begin{eqnarray}
  a_1 \equiv \frac{1}{\sqrt{2}}(\hat{x}^2+ i \hat{x}^1),
\quad
  a_2 \equiv \frac{1}{\sqrt{2}}(\hat{x}^4- i \hat{x}^3),
\end{eqnarray}
satisfying the standard commutation relation,
\begin{eqnarray}
 [a_i, a_j^\dagger ] =\delta_{ij},
\end{eqnarray}
and acting on the Fock space $\{ \ket{n,m} | n,m \geq 0, \in
{\bf Z}\}$.
Using this notation, the
$U(2)$ anti-self-dual instanton solution $X^\mu$ is expressed as
\begin{eqnarray}
  X^\mu &=& \Psi^\dagger \hat{x}^\mu \Psi \nn\\
&=&U^\dagger \hat{x}^\mu U
\label{solkf}
\\
&&+\rho^2
\left(
  \begin{array}{cc}
\left(2(N\!+\!2)\!+\!\rho^2\right)^{-1/2}
\hat{x}^\mu
\left(2(N\!+\!2)\!+\!\rho^2\right)^{-1/2}
& 0 \\
0 &
\left(2N\!+\!\rho^2\right)^{-1/2}
\hat{x}^\mu
\left(2N\!+\!\rho^2\right)^{-1/2}
  \end{array}
\right)
\nn
\end{eqnarray}
where
\begin{eqnarray}
  \Psi = \left(\Psi^{(1)}, \Psi^{(2)}\right),
\quad
\Psi^{(1)} = 
\left(  \begin{array}{c}
  \rho\\ 0\\ \sqrt{2}a_2^\dagger\\ -\sqrt{2}a_1^\dagger 
 \end{array}
\right)\frac{1}{\sqrt{2(N\!+\!2)\!+\!\rho^2}},
\quad
\Psi^{(2)} = 
\left(
  \begin{array}{c}
   0\\ \rho \\  \sqrt{2}a_1\\ \sqrt{2}a_2
  \end{array}
\right)\frac{1}{\sqrt{2N\!+\!\rho^2}}
\nn
\end{eqnarray}
and thus
\begin{eqnarray}
 U &\equiv& \left(N+2 + \frac{\rho^2}{2}\right)^{-1/2}
\left(
  \begin{array}{cc}
a_2^\dagger & a_1 \\ -a_1^\dagger & a_2
  \end{array}
\right).
\label{UU}
\end{eqnarray}
Here $N$ is the number operator $N\equiv a_1^\dagger a_1 + a_2^\dagger
a_2$ and $\rho$ is a parameter of the solution, which is related
the size of the solution as we will see below.
In the following, when we restore $\theta$,
we assign the dimension of length to the parameter $\rho$.

In the limit of $\rho \rightarrow 0$,
the solution (\ref{solkf}) becomes the zero size
instanton of the form (\ref{generalsolution}), as discussed in
\cite{KF}. To see this, we note that
the second term in (\ref{solkf})
disappears in this limit, and  the solution becomes
\begin{eqnarray}
  X^\mu = U_0^\dagger \hat{x}^\mu U_0
\label{four}
\end{eqnarray}
where the operator $U_0 \equiv U|_{\rho=0}$ satisfies\footnote{Note
  that, compared with the construction in the previous sections,
the roles of $U_0$ and $U_0^\dagger$ are exchanged.
In this section, we are following the notations of \cite{KF}.}
\begin{eqnarray}
  U_0U_0^\dagger=
\left(
  \begin{array}{cc}
\unit & 0 \\ 0 & \unit
  \end{array}
\right),
\quad
U_0^\dagger U_0=
\left(
  \begin{array}{cc}
\unit & 0 \\ 0 & \unit - \ket{0,0}\bra{0,0}
  \end{array}
\right).
\label{five}
\end{eqnarray}
Therefore this $U_0$ can be regarded as a shift operator, and
the Seiberg-Witten transform can be evaluated in the same way as in
the previous sections. For example, the D0-brane density is
given by
\begin{eqnarray}
  J(k) = {\bf Tr} e^{ik\cdot X} = 2\delta(k) + 1,
\label{510}
\end{eqnarray}
or in the $x$ space by
\begin{eqnarray}
  J(x) = {2 \over \theta^2} + \delta(x).
\end{eqnarray}
(Here we have restored $\theta$.)
The first term is for the uniform distribution of the D0-branes on the
parallel two D4-branes, and the second term gives the localized
additional D0-brane charge.

Now we consider the resolution of this
singularity by turning on the modulus $\rho$.
In the following,
we distinguish the three types of traces:
$\tr(\cdots)$ is over the $U(2)$ group indices,
${\rm Tr}(\cdots)$ is over the Fock space, and
the combined trace is expressed as
${\bf Tr} = \tr {\rm Tr}$.


\subsection{Small $\rho$ expansion}

Let us first turn on a small value of $\rho$ and see what happens.
The solution (\ref{solkf}) can be expanded in powers of $\rho$ as
\begin{eqnarray}
  \Psi^\dagger k\!\cdot\! \hat{x} \Psi
= A + \frac{\rho}{\sqrt{2}} (B+C) + \frac{\rho^2}{2}
 (D+E+F+G) + {\cal O}(\rho^3),
\end{eqnarray}
where
\begin{eqnarray}
&&  A \equiv U_0^\dagger k\!\cdot\! \hat{x} U_0, \\
&&B\equiv k\!\cdot\! \hat{x}\ket{0,0}\bra{0,0} \otimes
\left(
\begin{array}{cc}
0 & 0 \\
0 & 1
\end{array}
\right)
,
\quad C\equiv \ket{0,0}\bra{0,0} k\!\cdot\! \hat{x} \otimes
\left(
\begin{array}{cc}
0 & 0 \\
0 & 1
\end{array}
\right),\\
&&D \equiv \frac{1}{\sqrt{N_{\neq 0}}}k\! \cdot\! \hat{x}
 \frac{1}{\sqrt{N_{\neq 0}}} \otimes
\left(
\begin{array}{cc}
0 & 0 \\
0 & 1
\end{array}
\right),
\quad
E \equiv \frac{1}{\sqrt{N\!+\!2}}k\!\cdot\!  \hat{x}
 \frac{1}{\sqrt{N\!+\!2}}\otimes
\left(
\begin{array}{cc}
1 & 0 \\
0 & 0
\end{array}
\right),
\\
&&  F \equiv
-\Half U_0^\dagger \frac{1}{N\!+\!1}k\!\cdot\! \hat{x} U_0,
\quad  G \equiv -\Half U_0^\dagger k\!\cdot\! \hat{x}
\frac{1}{N\!+\!1} U_0.
\end{eqnarray}
The operator $1/N_{\neq 0}$ is defined in the projected Fock space
$\{(\unit-\ket{0,0}\bra{0,0})\ket{n,m}\}$.
Let us examine the D$0$-brane density of the solution
expanding again in powers of $\rho$,
\begin{eqnarray}
\lefteqn{ J(k)\equiv {\bf Tr}\left[\exp
    \left(
      i\Psi^\dagger  k\!\cdot\!\hat{x} \Psi
    \right)
\right]} \nn\\
&& ={\bf Tr}(e^{iA})
-\frac{\rho^2}{2} \int_0^1 d\tau{\bf Tr}
\left(
  Ce^{i\tau A}B
\right)
+\frac{\rho^2}{2}i
{\bf Tr}
\left(
  (D+E+F+G)e^{iA}
\right)
+{\cal O}(\rho^4).
\label{small}
\end{eqnarray}
Here we used relations
\begin{eqnarray}
  AC=BA=B^2 = C^2 =0.
\end{eqnarray}
As expected, the first term in the right-hand side of (\ref{small})
reproduces  (\ref{510}).
\begin{eqnarray}
 {\bf Tr}\left( e^{iA} \right)
 = 2\delta(k) + 1.
\label{ten}
\end{eqnarray}

Now we are going to evaluate the second term of the right-hand side
in (\ref{small}).
Using the relation (\ref{important}),
we obtain
\begin{eqnarray}
 {\bf Tr} [Ce^{i\tau k\cdot \hat{x}}B]
&=& \tr \bra{0,0}
k\!\cdot\!\hat{x}
\left(
  \begin{array}{cc}
0 & 0 \\ 0 & 1
  \end{array}
\right)
U_0^\dagger e^{i\tau k\cdot \hat{x}}U_0
\left(
  \begin{array}{cc}
0 & 0 \\ 0 & 1
  \end{array}
\right)
k\cdot\hat{x}\ket{0,0}\nn\\
&=&
{k^2 \over 2} \bra{0,0}e^{i\tau k\cdot \hat{x}}\ket{0,0 }\nn\\
&=&
{k^2 \over 2} e^{-\tau^2 k^2/4}.
\end{eqnarray}
Therefore the second term in (\ref{small}) can be written as
\begin{eqnarray}
-\frac{\rho^2}{2} \int_0^1 d\tau{\bf Tr}
\left(
  Ce^{i\tau A}B
\right)
=  -{1\over 4} k^2 \rho^2 \int_0^1 d\tau
e^{-\tau^2 k^2/4}.
\label{result1}
\end{eqnarray}

Let us proceed to the third term of the right-hand side of
(\ref{small}). First, we note
\begin{eqnarray}
  {\bf Tr}
[De^{iA}]
&=&
{\bf Tr}[U_0 D U_0^\dagger e^{ik\cdot\hat{x}}] \nn\\
&=&
{\rm Tr}
\left[
  \frac{1}{N+1}
(a_1\; k\!\cdot\!\hat{x}\;a_1^\dagger
+ a_2\; k\!\cdot\!\hat{x}\;a_2^\dagger)
  \frac{1}{N+1}
e^{ik\cdot\hat{x}}
\right]
\nn\\
&=&
{\rm Tr}
\left[
  \frac{1}{N+1}
k\!\cdot\!\hat{x}\frac{N+2}{N+1}e^{ik\cdot\hat{x}}
\right] \nn\\
&&~~~~~~+{1\over \sqrt{2}}{\rm Tr}
\left[
  \frac{1}{N+1}
\left((k_2 +ik_1)a_1^\dagger
+ (k_4-ik_3)a_2^\dagger\right)
  \frac{1}{N+1}
e^{ik\cdot\hat{x}}
\right],
\end{eqnarray}
where we have used the relation
(\ref{important}).
Similarly we can evaluate the other terms as
\begin{eqnarray}
&&  {\bf Tr}
[Ee^{iA}]
={\rm Tr}
\left[
  \frac{N}{N+1}
k\!\cdot\!\hat{x}\frac{1}{N+1}e^{ik\cdot\hat{x}}
\right] \nn\\
&&~~~~~~~~
-{1\over \sqrt{2}}{\rm Tr}
\left[
  \frac{1}{N+1}
\left((k_2 +ik_1)a_1^\dagger
+ (k_4-ik_3)a_2^\dagger\right)
  \frac{1}{N+1}
e^{ik\cdot\hat{x}}
\right],
\\
&&
{\bf Tr}[Fe^{iA}]  ={\bf Tr}[Ge^{iA}]  =
-
{\rm Tr}
\left[
\frac{1}{N+1}
k\!\cdot\!\hat{x} e^{ik\cdot\hat{x}}
\right].
\end{eqnarray}
Combining these together, we find that the third term
is actually zero. 
\begin{eqnarray}
{\bf Tr}
\left(
  (D+E+F+G)e^{iA}
\right)=0.
\label{result2}
\end{eqnarray}

Combining (\ref{ten}), (\ref{result1}), and (\ref{result2}),
the D0-brane density is given by
\begin{eqnarray}
  J(k)
=
2\delta^4(k) +1
-{1\over 4} \rho^2
k^2
\int_0^1 d\tau  \exp
\left(
  -\frac{k^2}{4}\tau^2
\right)
+ {\cal O}(\rho^4).
\label{eight}
\end{eqnarray}
Written in the $x$ representation by the Fourier transformation, the
D0-brane density is 
\begin{eqnarray}
 J(x)
= \frac{2}{\theta^2}+ \delta^4(x)
+
  \frac{\p^2}{\p x^i \p x^i}
{4\pi^2 \rho^2\over \theta^2}
\int_0^1 d\tau
\frac{1}{\tau^4}
\exp
\left(-
\frac{|x|^2}{\tau^2 \theta}
\right)
+{\cal O}
\left(
  {\rho^4 \over \theta^2}
\right)
\label{528}
\end{eqnarray}
Here we have restored $\theta$
using the dimensional analysis and the
convention that the parameter $\rho$
has the dimension of length.

Let us interpret this result.
The first term (\ref{528})  is for the uniformly
bounded D0-brane in the D4-brane, and the delta-function in
the second term represents the D0-brane of zero size.
Turning on $\rho$ deforms this delta-function singularity.
When $x \ll \sqrt{\theta}$, we can evaluate the
$\tau$-integral in the third term as
\begin{eqnarray}
  \frac{\p^2}{\p x^i \p x^i}
{4\pi^2 \rho^2\over \theta^2}
\int_0^1 d\tau
\frac{1}{\tau^4}
\exp
\left(-
\frac{|x|^2}{\tau^2 \theta}
\right)
=
 \rho^2\frac{\pi^{5/2}}{\sqrt{\theta}}
\frac{\p^2}{\p x^i\p x^i} \frac{1}{|x|^3} 
 + {\cal O}(1)
\end{eqnarray}
Therefore,  for $|x| \ll \sqrt{\theta}$, the D$0$-brane density of
the noncommutative instanton is
\begin{eqnarray}
J(x) - \frac{2}{\theta^2}
&= &   \delta^4(x) 
+ \rho^2\frac{\pi^{5/2}}{\sqrt{\theta}}
\frac{\p^2}{\p x^i\p x^i} \frac{1}{|x|^3} + \cdots \nn\\
&=& {-1 \over 2\pi^2}
 {\p^2 \over \p x^i \p x^i}
\left( {1\over |x|^2 } - \rho^2 {2\pi^{9/2} \over \sqrt{\theta}|x|^3} 
   \right) + \cdots,\quad (|x| \ll \sqrt{\theta}). 
\label{combi}
\end{eqnarray}
Thus the delta-function singularity in the $\rho=0$ solution 
is modified, suggesting that the singularity
is resolved for finite $\rho$. One can imagine,
for example, that
(\ref{combi}) represents the first two terms in the
$\rho$ expansion of the smooth function
\begin{eqnarray}
 \frac{\p^2}{\p x^i\p x^i} \frac{1}{(|x|+\rho^2/\sqrt{\theta})^2},
\label{rsolve}
\end{eqnarray}
where we neglected numerical coefficients. We will see in the
next subsection that, for large $\rho$, the D$0$-brane
density $J(x)$ indeed has a smooth profile. 

On the other hand, for $|x| \gg \sqrt{\theta}$, 
the $\tau$-integral in (\ref{528}) can also be evaluated
and the D$0$-brane density is given by
\begin{eqnarray}
J(x) - {2 \over \theta^2} = \frac{2\pi^2\rho^2}{\theta}
   \frac{\p^2}{\p x^i\p x^i} 
\left[\frac{1}{|x|^2}\exp
   \left(
     -\frac{|x|^2}{\theta}
   \right)
\right] + \cdots, \quad (|x| \gg \sqrt{\theta}).
\end{eqnarray}
Thus the asymptotic behavior of the D0-brane charge distribution is
Gaussian with the width $\sim \sqrt{\theta}$.

The $U(1)$ part of the field strength, $i.e.$, the D$2$-brane
density, can be evaluated
in a similar fashion.
Using the expansion
\begin{eqnarray}
  [\Psi^\dagger a_1 \Psi, \Psi^\dagger a_1^\dagger \Psi]
= U_0^\dagger U_0 + \frac{\rho^2}{2}
\left(
  \begin{array}{cc}
\frac{1}{(N+1)(N+2)} & 0 \\
0 & \ket{0,0}\bra{0,0}-\frac{1}{N_{\neq 0}(N+2)}
  \end{array}
\right)
+{\cal O}(\rho^4),
\end{eqnarray}
we have
\begin{eqnarray}
&&
{\bf Tr}  \left[
[\Psi^\dagger a_1 \Psi, \Psi^\dagger a_1^\dagger \Psi]
e^{ik\cdot X}
\right]\\
\label{expxx}
&& =
{\bf Tr}
\left[
  U_0^\dagger U_0 e^{ik\cdot X}
\right]
+\rho^2
{\bf Tr}
\left[
\left(
  \begin{array}{cc}
\frac{1}{(N+1)(N+2)} & 0 \\
0 & \ket{0,0}\bra{0,0}-\frac{1}{N_{\neq 0}(N+1)}
  \end{array}
\right)
e^{iA}
\right]
+{\cal O} (\rho^4).
\nn
\end{eqnarray}
The first term in the right-hand side is evaluated in the same fashion,
and the result is
\begin{eqnarray}
\lefteqn{{\bf Tr}
\left[
  U_0^\dagger U_0 e^{ik\cdot X}
\right]}\nn\\
&&
=
\delta^4(k)
+ \frac{\rho^2}{2}
\left[
-  \int_0^1 d\tau {\rm Tr}
  \left[
    C e^{i\tau A}B
  \right]
+  \int_0^1 d\tau' \tau'\int_0^1 d\tau {\rm Tr}
  \left[
    C e^{i\tau\tau' A}B
  \right]
\right]
+{\cal O}(\rho^4)\nn\\
&&
=
\delta^4(k)
+ \frac{\rho^2}{2}
\left[
-  \int_0^1 d\tau
\frac{|k|^2}{2}
e^{-\tau^2|k|^2/4}
+  \int_0^1 d\tau' \tau'\int_0^1 d\tau
\frac{|k|^2}{2}
e^{-\tau^2{\tau'}^2|k|^2/4}
\right]
+{\cal O}(\rho^4).
\label{integ}
\end{eqnarray}
The second term in (\ref{expxx}) is turned out to be simple,
\begin{eqnarray}
  \frac{\rho^2}{2}
  \left(
    1-e^{-|k|^2/4}
  \right).
\end{eqnarray}
Summing up all the contributions and noting that the second integral
in (\ref{integ}) is arranged to cancel with the error function coming
from the first integral, we found that the result vanishes:
\begin{eqnarray}
{\bf Tr}  \left[
[\Psi^\dagger a_1 \Psi, \Psi^\dagger a_1^\dagger \Psi]
e^{ik\cdot X}
\right]
=\delta^4(k) + 0 + {\cal O}(\rho^4).
\end{eqnarray}
Therefore, the Seiberg-Witten transform of the $U(1)$
part of the field strength vanishes
\begin{eqnarray}
\tr F_{34} (x)
=0+ {\cal O}(\rho^4).
\label{u1part}
\end{eqnarray}
Similarly one can show that all other components
vanish to this order,
\begin{eqnarray}
\tr F_{ij}(x)= 0 + {\cal O}(\rho^4).
\label{eleven}
\end{eqnarray}

In fact one can show that, if ${\rm tr} F_{ij}$ is smooth and
decays sufficiently fast at the infinity, it vanishes
identically,
\begin{eqnarray}
  {\rm tr}F_{ij} = 0.
\end{eqnarray}
To see this, we note that the anti-self-dual equation,
\begin{eqnarray}
  [X^i, X^j] = -{1\over 2}\epsilon_{ijkl} [X^k, X^l],
\end{eqnarray}
implies, via the Seiberg-Witten map, that
${\rm tr} F_{ij}$ is also anti-self-dual.
Since ${\rm tr} F_{ij}$ obeys the Bianchi identity
as shown in \cite{OO},
we can write ${\rm tr} F_{ij} = \partial_{[i,}
a_{j]}$ for some $U(1)$ gauge field $a_i$.
It is well-known that there is no non-trivial
solution to the anti-self-dual equation in
the $U(1)$ gauge theory. Thus it should vanish
identically for any $\rho$, assuming it is smooth
and vanish sufficiently fast for large $x$. 
One can also argue that the BPS instanton solution
considered here should not carry any local D$2$-brane charges. 
The computation at large $\rho$, in the next subsection,
also shows that ${\rm tr} F_{ij}$ vanishes. 

\subsection{Large $\rho$ expansion}

Before going into detailed calculation of the large $\rho$
expansion, let us take a look at the limit $\rho = \infty$.
There we have
\begin{eqnarray}
  X^\mu = \hat{x}^\mu \unit_{2\times 2}.
\end{eqnarray}
Note that the non-zero contribution is coming from the second term of
the solution (\ref{solkf}), not from the first term, which dominates
in the opposite limit $\rho=0$. It is clear that
the Seiberg-Witten map gives zero
gauge field and vanishing D0-brane density. This is
consistent with the expectation that, in the large $\rho$ limit, the
instanton spreads over and the structure of the soliton disappears.

Now let us evaluate the sub-leading terms in the $1/\rho$ expansion,
\begin{eqnarray}
  \Psi^\dagger k\!\cdot\! \hat{x}\Psi
= k\!\cdot\! \hat{x} + \frac{2}{\rho^2}P +
\frac{4}{\rho^4}Q +
\frac{8}{\rho^6}R +
{\cal O}
\left(
  1/\rho^8
\right),
\label{expandl}
\end{eqnarray}
where
\begin{eqnarray}
&&  P \equiv\Half k\!\cdot\! \hat{x}
\otimes
\left(
\begin{array}{cc}
1 & 0 \\ 0 & -1
\end{array}
\right)
+
\left(
\begin{array}{cc}
0 & p_2 a_1-p_1 a_2 \\
(p_2 a_1 - p_1 a_2)^\dagger & 0
\end{array}
  \right), \label{firstcomplex}\\
&&
Q \equiv
\frac{-3}{8}k\!\cdot\! \hat{x}
\otimes\unit_{2\times 2}
-\Half P (N\!+\!1) -\Half (N\!+\!1) P,
\\
&& \tr R =
\frac34
\left(Nk\!\cdot\! \hat{x} + k\!\cdot\! \hat{x} N
+ 2 k\!\cdot\! \hat{x}\right).
\label{secondcomplex}
\end{eqnarray}
In (\ref{firstcomplex}), we used the complex combination of
the momentum $k$ defined as
\begin{eqnarray} p_1 = {1\over \sqrt{2}}
(k_2 + i k_1),~~~p_2 = {1\over \sqrt{2}}
(k_4 - i k_3).
\label{complexmom}
\end{eqnarray}
We did not write down the explicit form of $R$ since
only its $U(2)$ trace,
$\tr R$, is going to be necessary in the following.
To evaluate $Q$ and $\tr R$, we have used the relation
\begin{eqnarray}
  [N, [N, k\!\cdot\! \hat{x}]] = k\!\cdot\! \hat{x}.
\label{usefull}
\end{eqnarray}

Let us compute the D0-brane density
\begin{eqnarray}
 J(k)\equiv {\bf Tr} \left[\exp
    \left(
      i\Psi^\dagger k\!\cdot\!\hat{x}\Psi
    \right)
\right].
\end{eqnarray}
It turns out that the ${\cal O}(\rho^{-2})$ term vanishes since
$\tr P = 0$. Thus
we have to start with the ${\cal O}(\rho^{-4})$ terms.

Using the cyclic property of the trace ${\bf Tr}$, we
find
\begin{eqnarray}
\frac14 J(k)\biggm|_{{\rm order}(1/\rho^4)}
&=&
{\bf Tr}
\left[
iQ  e^{ik\cdot \hat{x}}
\right]
+{\bf Tr}
\sum_{n=0}^{\infty}\sum_{l,m\geq 0}
\left[
(ik\!\cdot\! \hat{x})^l
iP (ik\!\cdot\! \hat{x})^m
iP (ik\!\cdot\! \hat{x})^{n-2-l-m}
\right] \nn\\
&=&
{\bf Tr}
\left[
iQ  e^{ik\cdot \hat{x}}
\right]
+
{\bf Tr}
\left[
(iP)^2  e^{ik \hat{x}}
\right],
\label{pq2}
\end{eqnarray}
where we used the fact that $P$ and $k\!\cdot\! \hat{x}$ commute.
To evaluate the traces, we employ the following
formulae proven in Appendix A,
\begin{eqnarray}
a_1 e^{ik\cdot\hat{x}} = -i
\left(
  \frac{\p}{\p\bar{p}_1}
-\Half p_1
\right)e^{ik\cdot\hat{x}}, ~~~~etc,
\label{eq2}
\end{eqnarray}
where $p_1, p_2$ are the complex combination
of the momentum (\ref{complexmom}).
The result is
\begin{eqnarray}
\frac14 J(k)\biggm|_{{\rm order} 1/\rho^4}
&=&  8\delta^4(k)
+
\left(
  |p_1|^2\frac{\p}{\p p_2}\frac{\p}{\p \bar{p}_2}
+
  |p_2|^2\frac{\p}{\p p_1}\frac{\p}{\p \bar{p}_1}\right.\nn\\
&&\hspace{30mm}\left. -
  \bar{p}_1 p_2\frac{\p}{\p p_2}\frac{\p}{\p \bar{p}_1}
-
  \bar{p}_2p_1 \frac{\p}{\p p_1}\frac{\p}{\p \bar{p}_2}
\right)\delta^4(k).
\end{eqnarray}
This is further simplified by
\begin{eqnarray}
p_1 \frac{\p}{\p p_1}   \delta^4(k) = - \delta^4(k),
\end{eqnarray}
and finally we obtain
\begin{eqnarray}
  J(k)\biggm|_{{\rm order} 1/\rho^4} =  \frac{24}{\rho^4} \delta^4(k).
\end{eqnarray}
Therefore, in terms of the commutative $x$ coordinates,
the ${\cal O}(\rho^{-4})$ term in the D$0$-brane density is
\begin{eqnarray}
  J(x)\biggm|_{{\rm order} 1/\rho^4} =  \frac{24}{\rho^4}.
\label{result}
\end{eqnarray}
Remarkably, this agrees with the $1/\rho$ expansion
of the BPST instanton in the commutative gauge theory:
\begin{eqnarray}
  {}F_{\mu\nu} = \frac{4\rho^2}{(|x|^2 + \rho^2)^2} \Sigma_{\mu\nu}
\end{eqnarray}
where $\Sigma_{\mu\nu} \equiv \eta^{i\mu\nu}\sigma_i$ with
the Pauli matrix $\sigma_i (i=1,2,3)$
and the 'tHooft symbol $\eta$.
Substituting this into the D$0$-brane density
\begin{eqnarray}
\frac18 \tr \epsilon^{ijkl}(F_{ij} - \theta^{-1}_{ij})
(F_{kl} - \theta^{-1}_{kl})
\label{FF}
\end{eqnarray}
and expand it in powers of $1/\rho$,
we find
\begin{eqnarray}
\lefteqn{-\frac1{8}
  \tr \epsilon^{ijkl}(F_{ij} - \theta^{-1}_{ij}\unit)
(F_{kl} - \theta^{-1}_{kl}\unit)}\nn\\
&=&\frac{2}{\theta^2} +
\frac1{8}
\tr \epsilon^{ijkl}F_{ij}F_{kl}\nn\\
&=& \frac{2}{\theta^2} +
\frac{24}{\rho^4} - {96 \over \rho^6} |x|^2 +  {\cal O}
\left(
  \frac1{\rho^8}
\right).
\label{coin}
\end{eqnarray}
The ${\cal O}(\rho^{-4})$ term
exactly agrees with the above calculation (\ref{result}).

The fact that the noncommutative instanton becomes
the commutative one in the limit $\theta \rightarrow 0$
does not by itself guarantees this agreement.
For example, there could have been a correction
of the form $e^{-x^2/\theta}$ multiplying $\rho^{-4}$, 
which vanishes in the commutative limit. 
Such a correction is absent since 
the structure of the expansion given 
by (\ref{firstcomplex}) - (\ref{secondcomplex})
suggests that the coefficients of the $1/\rho$ expansion
are polynomials in $x$. By a simple dimensional
analysis, one can show that, under this condition,
no $\theta$ dependent term is allowed in the $0(\rho^{-4})$ order.
Therefore the agreement of the number $24$ gives
a nice consistency check of our computation.

We have gone further and carried out the ${\cal O}(\rho^{-6})$
computation of the D$0$-brane density.
The detail is given in Appendix B. The result is even more
surprising:
\begin{eqnarray}
  J(x)\biggm|_{{\rm order} 1/\rho^6} =
- \frac{96}{\rho^6}
|x|^2.
\label{surprise}
\end{eqnarray}
This term perfectly agrees with the corresponding
term in (\ref{coin}). Thus, even to this order,
there is no corrections to the D$0$-brane
distribution due to the noncommutativity.
We should point out that, to this order,
there could have been a term of the form
$\theta/\rho^6$, but the coefficient
in front of it turned out to be zero.

We have also computed the $U(1)$ part of the
field strength, $i.e.$, the D$2$-brane density.
The leading term is of the order ${\cal O}(\rho^{-2})$,
but it turned out to be zero, in agreement
with expectation that the BPS instanton does not carry any D$2$
brane charge.


\section{Conclusion}

In this paper, we have evaluated the Seiberg-Witten map
for various solitons and instantons in noncommutative gauge
theory. When the gauge theory is defined by the low energy
limit of string theory, the Seiberg-Witten map
describes how these solutions couple to the Ramond-Ramond
potentials of closed string theory \cite{OO,Mukhi,Jeremy}.
Therefore, by studying the Seiberg-Witten map, we can read
off various information about Ramond-Ramond charge distributions
of these solutions.

We find that the Ramond-Ramond charge distributions of
solutions constructed using projection operators
have delta-function supports. They include solutions in
two-dimensional Yang-Mills theory (\ref{generalsolution}), pure
D$0$-brane in various dimensions  (\ref{obvious}), intersecting
D$2$-branes
(\ref{orthogonal}),
(\ref{additionalmoduli2})-(\ref{additionalmoduli}),
(\ref{add1})-(\ref{add2}), and
D$1$-branes
intersecting with
D$3$-brane (\ref{fluxonsol}).

On the other hand, instantons in higher dimensions allow
deformation away from the projection operator construction
and therefore their Seiberg-Witten transforms can have
finite sizes. We studied in detail the case of the $U(2)$
anti-self-dual instanton given by (\ref{solkf})-(\ref{UU}). The solution
has the deformation parameter $\rho$. In the limit $\rho
\rightarrow 0$, the solution reduces to the one for
the pure D$0$-brane (\ref{obvious}).
Turning on a small amount of
$\rho$, the D$0$-brane density is deformed as in
(\ref{528}). We see that the D$0$-brane charge is now distributed
over the region of size $\sim \sqrt{\theta}$. In
addition, the delta-function singularity of the D$0$-brane charge
distribution is modified as
\begin{eqnarray}
\delta(x) = \frac{-1}{2\pi^2} {\p^2 \over 
\p x^i \p x^i} {1 \over |x|^2} \rightarrow
{-1 \over 2\pi^2}
 {\p^2 \over \p x^i \p x^i}
\left( {1\over |x|^2 } - \rho^2 {2\pi^{9/2} \over \sqrt{\theta}|x|^3}
   \right).
\end{eqnarray}

{}For large $\rho$, we can evaluate the Seiberg-Witten map
of the instanton in the $1/\rho$ expansion. We find that
the D$0$-brane density of the noncommutative instanton
agrees surprisingly well with that of the commutative
instanton. The agreement in the leading terms, (\ref{result})
and (\ref{coin}), is expected and gives a nice consistency
check of our computation. The agreement of the sub-leading
term, (\ref{surprise}) and (\ref{coin}), is surprising and we do not
have an explanation of this phenomenon.

We also find that the $U(1)$ part of the Seiberg-Witten map
vanishes for both small $\rho$ and large $\rho$.
Since there is no nontrivial
anti-self-dual solution in the $U(1)$ gauge theory in
commutative space, we expect that ${\rm tr} F_{ij}$
vanishes for any $\rho$. It is consistent with 
the expectation that the BPS instanton should
not carry any local D$2$-brane charges.

In \cite{David} -- \cite{GH}, Seiberg-Witten transform of
noncommutative monopoles are studied with fixed $\alpha'$ and small
$\theta$. This is in contrast to our case where we
use the exact Seiberg-Witten map of \cite{OO,Mukhi,Jeremy} 
in the Seiberg-Witten limit ($\alpha' \rightarrow 0$)
and with finite $\theta$. It will be interesting to extend
this analysis to include the case studied
in \cite{David} - \cite{GH}. 

In this paper, we have evaluated the Seiberg-Witten
map for the $U(1)$ part of the field strength.
It is desirable to find an explicit expression for
the non-Abelian part of the Seiberg-Witten map since
it would carry more information on these solutions.
Progress in this direction has been made in
\cite{wess,Zumino}. (For our purpose, we need an
inverse of the map studied in these papers.)


\vs{10mm}
\noindent
{\large \bf Acknowledgments}

We thank Yuji Okawa for useful discussions
and for comments on the earlier version of this paper.
H.O. thanks the Institute for Theoretical Physics,
Santa Barbara, for the hospitality.

K. H. was supported in part
by Japan Society for the Promotion of
Science under the Postdoctoral Research Program (\# 02482).
H. O. was  supported in part by the Department of Energy grant
DE-FG03-92ER40701 and the Caltech Discovery Fund.
In addition, this research was supported in part by the National
Science Foundation under Grant No.\ PHY99-07949.


\appendix


\section{Useful formulae}
In this appendix we derive the formula (\ref{eq2}) and other
useful formulae used in the evaluation of the large $\rho$ expansion
in Sec.\ 5.2.
We find it useful to introduce the complex combinations of
the momentum $k$ as
\begin{eqnarray} p_1 = {1\over \sqrt{2}}
(k_2 + i k_1),~~~p_2 = {1\over \sqrt{2}}
(k_4 - i k_3),
\end{eqnarray}
so that the following relation holds:
\begin{eqnarray}
  k\!\cdot\!\hat{x} =
p_1 a_1^\dagger + \bar{p} a_1 + p_2 a_2^\dagger +
\bar{p}_2 a_2.
\end{eqnarray}

To show (\ref{eq2}) is easy, by acting a derivative on
$e^{ik\cdot\hat{x}}$ as
\begin{eqnarray}
  \frac{\p}{\p \bar{p}_1}e^{i k\cdot\hat{x}}
=&&\sum_n \frac{(i)^n}{n!}
\sum_{m=0}^{n-1}
(k\!\cdot\! \hat{x})^m a_1 (k\!\cdot\! \hat{x})^{n-1-m}\nn\\
=&&\sum_n \frac{(i)^n}{n!}
\left(
n a_1 (k\!\cdot\! \hat{x})^{n-1}
+
\sum_{m=0}^{n-1}
(-p_1)(k\!\cdot\! \hat{x})^{n-2}
\right)
\nn\\
=&&ia_1 e^{ik\cdot \hat{x}} + \Half p_1 e^{ik\cdot\hat{x}}.
\end{eqnarray}
This verifies (\ref{eq2}).

In the following, we shall derive a useful formula which is necessary
in evaluating the $1/\rho^6$ contribution in the D0-brane density in
Appendix B.
{}For simplicity
we consider 2 dimensional noncommutative space and evaluate
\begin{eqnarray}
  {\rm Tr}[n e^{ik\cdot\hat{x}}].
\end{eqnarray}
Acting the derivative twice, we easily obtain
\begin{eqnarray}
  {\rm Tr}[a^\dagger a e^{ik\cdot\hat{x}}]
=-i
\left(
  \frac{\p}{\p \bar{p}} - \Half p
\right)
\left[-i
\left(
  \frac{\p}{\p p} + \Half \bar{p}
\right)
\delta^2(k)
\right].
\end{eqnarray}
Here note the order of the differentiation. Taking care of the formula
\begin{eqnarray}
x \p_x \delta(x) = -\delta(x),
\end{eqnarray}
we obtain
\begin{eqnarray}
  {\rm Tr}[a^\dagger a e^{ik\cdot\hat{x}}]
=
\left(
-\Half-\frac{\p}{\p p}\frac{\p}{\p \bar{p}}
\right)
\delta^2(k).
\end{eqnarray}
Therefore, for $N\equiv a_1^\dagger a_1 + a_2^\dagger a_2$,
we obtain
\begin{eqnarray}
  {\rm Tr} [N e^{ik\cdot\hat{x}}] =
\left(
-1-\frac{\p}{\p p_1}\frac{\p}{\p \bar{p}_1}
-\frac{\p}{\p p_2}\frac{\p}{\p \bar{p}_2}
\right)\delta^4(k).
\label{eq1}
\end{eqnarray}


\section{Evaluation of Order $\rho^{-6}$ terms in the $U(2)$ Instanton}

In this appendix, we derive the sub-leading result (\ref{surprise}).

The contribution of this order ${\cal O}(1/\rho^6)$ in the D0-brane
current density $\exp[i\Psi^\dagger k\!\cdot\! \hat{x}\Psi]$ is
\begin{eqnarray}
&&8\sum_n\frac{i^n}{n!}
\left(
\sum_{m_1,m_2,m_3\geq 0}  {\bf Tr}
(k\!\cdot\! \hat{x})^{m_1} P (k\!\cdot\! \hat{x})^{m_2} P
(k\!\cdot\! \hat{x})^{m_3} P (k\!\cdot\! \hat{x})^{n-3-m_1-m_2-m_3}
\right.
\nn\\
&& \hs{20mm}
+\sum_{m_1,m_2\geq 0}  {\bf Tr}
(k\!\cdot\! \hat{x})^{m_1} P (k\!\cdot\! \hat{x})^{m_2} Q
(k\!\cdot\! \hat{x})^{n-2-m_1-m_2}
\nn\\
&& \hs{35mm}
+\sum_{m_1,m_2\geq 0}  {\bf Tr}
(k\!\cdot\! \hat{x})^{m_1} Q (k\!\cdot\! \hat{x})^{m_2} P
(k\!\cdot\! \hat{x})^{n-2-m_1-m_2}
\nn\\
&& \hs{50mm}
\left.
+\sum_{m\geq 0}  {\bf Tr}
(k\!\cdot\! \hat{x})^m R (k\!\cdot\! \hat{x})^{n-1-m}
\right).
\label{lastt}
\end{eqnarray}
Using the cyclic property of the trace under that these
summation over $n$ can be expressed in terms of the compact operator
$e^{ik\cdot\hat{x}}$, we can rewrite this
as\footnote{For example, the
  last term in (\ref{lastt}) is rearranged without using the
  cyclicity as
  \begin{eqnarray}
    \int_0^1d\tau \; {\bf Tr}\left[ e^{i\tau k \cdot\hat{x}} R
     e^{i(1-\tau) k \cdot\hat{x}}\right].
  \end{eqnarray}
However, concerning the first term in (\ref{lastt}), it is not
necessary to use the cyclic property because $P$ is commutative with
$k\!\cdot\! \hat{x}$. }
\begin{eqnarray}
 8 {\bf Tr}
  \left[
    (iP)^3e^{ik\cdot\hat{x}}
   +(iQ)(iP)e^{ik\cdot\hat{x}}
   +iRe^{ik\cdot\hat{x}}
  \right].
\label{tbe}
\end{eqnarray}
Let us evaluate each term in the trace respectively.

The first term turns out to be vanishing. This is because
\begin{eqnarray}
  \tr[P^3] = \frac14 (k\!\cdot\! \hat{x}) k^2
\end{eqnarray}
and thus
\begin{eqnarray}
{\bf Tr}
\left[
  (iP)^3 e^{ik\cdot\hat{x}}
\right]
=
 -\frac14 k^2 {\rm Tr} \left[
ik\!\cdot\! \hat{x} e^{ik\cdot\hat{x}}
\right]
=k^2\delta^4(k) =0.
\label{rer1}
\end{eqnarray}

The second term in (\ref{tbe}) is calculated in the
following. First, taking the $U(2)$ trace, we have
\begin{eqnarray}
  {\bf Tr}
  \left[
    QPe^{ik\cdot\hat{x}}
  \right]
=-
{\rm Tr}
\left[
  (N+1)
  \left(
    \Half (k\!\cdot\! \hat{x})^2 + \Half k^2+ 2(\bar{p}_2 a_1^\dagger
- \bar{p}_1 a_2^\dagger) (p_2 a_1 - p_1 a_2)
  \right)e^{ik\cdot\hat{x}}
\right].
\end{eqnarray}
Using the formula (\ref{eq1}), the first term of this expression is
evaluated as
\begin{eqnarray}
{\rm Tr}
\left[
  (N+1)
    \Half (k\!\cdot\! \hat{x})^2 e^{ik\cdot\hat{x}}
\right]
& = &
-\Half
\left(
  \frac{\p}{\p \tau}
\right)^2
{\rm Tr} \left[(N+1) e^{i\tau k\cdot\hat{x}}\right]
\biggm|_{\tau=1}
\nn\\
&=&
-\Half
\left(
  \frac{\p}{\p \tau}
\right)^2
\left[
  \left(
-    \frac{\p}{\p(\tau p_1)}    \frac{\p}{\p(\tau \bar{p}_1)}
-    \frac{\p}{\p(\tau p_2)}    \frac{\p}{\p(\tau \bar{p}_2)}
  \right)
\delta^4(\tau k)
\right]
\biggm|_{\tau=1}
\nn\\
&=&
21
  \left(
    \frac{\p}{\p p_1}    \frac{\p}{\p \bar{p}_1}
+    \frac{\p}{\p p_2}    \frac{\p}{\p \bar{p}_2}
  \right)
\delta^4(k).
\end{eqnarray}
We calculate the second term in the similar way and obtain
\begin{eqnarray}
  {\rm Tr}
\left[
 \Half (N+1)
k^2 e^{ik\cdot\hat{x}}
\right]
=-2\delta^4(k).
\end{eqnarray}
The third term is slightly complicated, however using the formula
(\ref{eq2}) and (\ref{eq1}) the straightforward calculation shows
\begin{eqnarray}
\lefteqn{
{\rm Tr}
\left[
  (N+1)
  \left(
(\bar{p}_2 a_1^\dagger
- \bar{p}_1 a_2^\dagger) (p_2 a_1 - p_1 a_2)
  \right)e^{ik\cdot\hat{x}}
\right]}
\nn\\
&&
=
\left[
-\bar{p}_1 p_1
  \left(
    \frac{\p}{\p \bar{p}_2}-\Half p_2
  \right)
  \left(
    \frac{\p}{\p p_2}+\Half \bar{p}_2
  \right)
-\bar{p}_2 p_2
  \left(
    \frac{\p}{\p \bar{p}_1}-\Half p_1
  \right)
  \left(
    \frac{\p}{\p p_1}+\Half \bar{p}_1
  \right)
\right.
\nn\\
&& \hs{10mm}
\left.
+\bar{p}_2 p_1
\left(
  \frac{\p}{\p \bar{p}_2} - \Half p_2
\right)
\left(
  \frac{\p}{\p p_1} + \Half \bar{p}_1
\right)
+\bar{p}_1 p_2
\left(
  \frac{\p}{\p \bar{p}_1} - \Half p_1
\right)
\left(
  \frac{\p}{\p p_2} + \Half \bar{p}_2
\right)
\right]
\nn\\
&&\hs{10mm}
\times
\left(
  -\frac{\p}{\p p_1}\frac{\p}{\p \bar{p}_1}
  -\frac{\p}{\p p_2}\frac{\p}{\p \bar{p}_2}
\right)
\delta^4(k)
\nn\\
&& =
\left(
  1
  -3\frac{\p}{\p p_1}\frac{\p}{\p \bar{p}_1}
  -3\frac{\p}{\p p_2}\frac{\p}{\p \bar{p}_2}
\right)
\delta^4(k).
\end{eqnarray}
Therefore, summarizing them, we have
\begin{eqnarray}
  {\bf Tr}(iQ)(iP)e^{ik\cdot\hat{x}}
=
15\left(
\frac{\p}{\p p_1}\frac{\p}{\p \bar{p}_1}
+\frac{\p}{\p p_2}\frac{\p}{\p \bar{p}_2}
\right)
\delta^4(k).
\label{rer2}
\end{eqnarray}

The third term in (\ref{tbe}) is rather easily evaluated by using
the formula (\ref{eq1}), and the result is
\begin{eqnarray}
    {\bf Tr}(iR)e^{ik\cdot\hat{x}}
=
9\left(
\frac{\p}{\p p_1}\frac{\p}{\p \bar{p}_1}
+\frac{\p}{\p p_2}\frac{\p}{\p \bar{p}_2}
\right)
\delta^4(k).
\label{rer3}
\end{eqnarray}

Summing up all the contribution (\ref{rer1}), (\ref{rer2}) and
(\ref{rer3}), we obtain the order $1/\rho^6$ result as
\begin{eqnarray}
  \frac{192}{\rho^6}
\left(
\frac{\p}{\p p_1}\frac{\p}{\p \bar{p}_1}
+\frac{\p}{\p p_2}\frac{\p}{\p \bar{p}_2}
\right)
\delta^4(k).
\end{eqnarray}
Restoring the $\theta$ dependence and noting the relations
\begin{eqnarray}
  \frac{\p}{\p p_1}\frac{\p}{\p \bar{p}_1}
=
\frac{1}{2}
\left[
\frac{\p^2}{\p k_1 \p k_1}
+\frac{\p^2}{\p k_2\p k_2}
\right],
\end{eqnarray}
we obtain
\begin{eqnarray}
 J(k)\biggm|_{{\rm order} (\theta^3/\rho^6)}
= \frac{96}{\rho^6}
  \frac{\p^2}{\p k_i \p k_i}
\delta^4(k).
\end{eqnarray}
Performing the Fourier transformation, we obtain the result
(\ref{surprise}).


\newcommand{\J}[4]{{\sl #1} {\bf #2} (#3) #4}
\newcommand{\andJ}[3]{{\bf #1} (#2) #3}
\newcommand{\AP}{Ann.\ Phys.\ (N.Y.)}
\newcommand{\MPL}{Mod.\ Phys.\ Lett.}
\newcommand{\NP}{Nucl.\ Phys.}
\newcommand{\PL}{Phys.\ Lett.}
\newcommand{\PR}{ Phys.\ Rev.}
\newcommand{\PRL}{Phys.\ Rev.\ Lett.}
\newcommand{\PTP}{Prog.\ Theor.\ Phys.}

\end{document}